\documentclass[aps,apl,twocolumn,showpacs,superscriptaddress]{revtex4}

\usepackage{graphicx} 
\usepackage{soul}
\begin{document}

\title{ Magnetization of Graphane by Dehydrogenation }

\author{H. \c{S}ahin}
\affiliation{UNAM-Institute of Materials Science and
Nanotechnology, Bilkent University, 06800 Ankara, Turkey}

\author{C. Ataca}
\affiliation{UNAM-Institute of Materials Science and
Nanotechnology, Bilkent University, 06800 Ankara, Turkey}
\affiliation{Department of Physics, Bilkent University, 06800 Ankara,
Turkey}

\author{S. Ciraci}\email{ciraci@fen.bilkent.edu.tr}
\affiliation{UNAM-Institute of Materials Science and
Nanotechnology, Bilkent University, 06800 Ankara, Turkey}
\affiliation{Department of Physics, Bilkent University, 06800 Ankara, Turkey}

\date{\today}

\begin{abstract}

Using first principles calculations, we show that each hydrogen
vacancy created at graphane surface results in a local unpaired
spin. For domains of hydrogen vacancies the situation is, however
complex and depends on the size and geometry of domains, as well as
whether the domains are single- or double-sided. In single-sided
domains, hydrogen atoms at the other side are relocated to pair the
spins of adjacent carbon atoms by forming $\pi$-bonds. Owing to the
different characters of exchange coupling in different ranges and
interplay between unpaired spin and the binding geometry of
hydrogen, vacancy domains can attain sizable net magnetic moments.
\end{abstract}


\maketitle

Graphene\cite{novo}, a truly two-dimensional (2D) crystal of
honeycomb structure, has sparked considerable interest not only
because of its charge carriers behaving like massless Dirac
fermions\cite{novo2, zhang, peres}, but also the unusual magnetic
properties displayed by its flakes and
nanoribbons\cite{nanoribbon,haldun,engin,can,hasan,dot}. In
addition to numerous experimental and theoretical studies on the
physical properties of graphene, efforts have been also devoted to
synthesize various types of derivatives of graphene. More recently,
a 2D hydrocarbon material in the family of honeycomb structure,
namely \textit{graphane} is synthesized\cite{novo-graphane}.
Interesting properties such as reversible
hydrogenation-dehydrogenation with changing
temperature\cite{novo-graphane}, the electronic structure
with a wide band gap\cite{sofo, boukhvalov} have been revealed
soon after its synthesis. In this letter, we reveal that graphane
can be magnetized by dehydrogenation of domains on its surfaces.
Large magnetic moments can be attained in a small domain on the
 graphane sheet, depending on whether the defect region is
one-sided or two-sided. Our predictions are obtained from the
state-of-the-art spin polarized first-principles plane-wave
calculations\cite{dft,gw} within the LDA noncollinear calculations including spin-orbit interaction, using (11$\times$11$\times$1) supercells. Details of our method can be found in \cite{hasan-ansiklopedi, nc}.

\begin{figure}
\includegraphics[width=8.5cm]{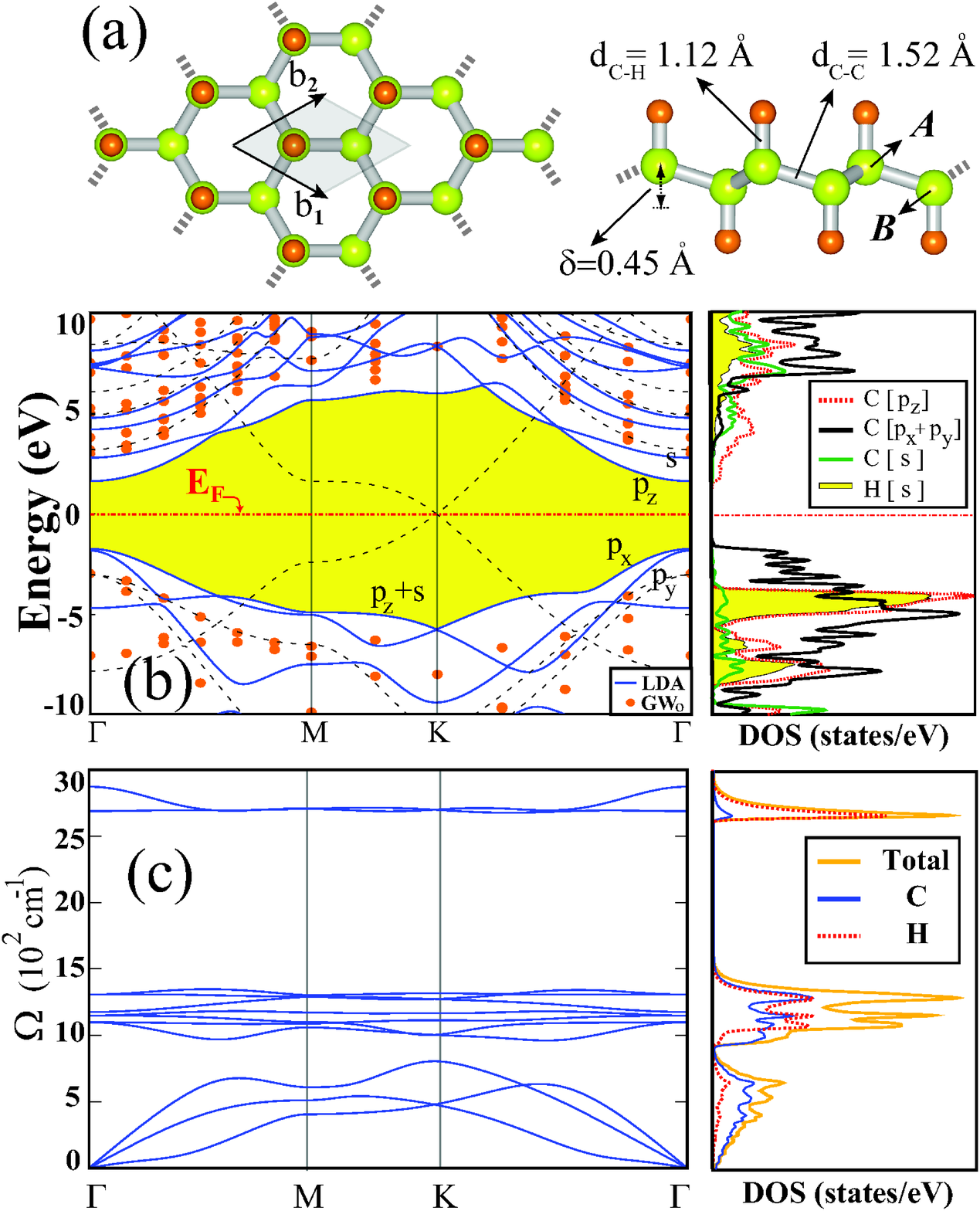}
\caption{(Color online) (a) Top and side views of atomic structure
showing of graphane primitive unit cell with Bravais lattice vectors
\textbf{b$_1$} and \textbf{b$_2$} and buckling of alternating
carbon atoms, A and B, in honeycomb structure $\delta$, bond lengths
$d_{C-C}$ and $d_{C-H}$ optimized using LDA. Large green (light)
and small orange (dark) balls indicate C and H atoms, respectively.
(b) Energy band structure is calculated by using LDA and corrected using GW$_0$ (shown by blue lines and orange dots). For graphene, linear band crossing at Dirac point is shown by dashed grey lines.
(c) Calculated phonon bands and density of states DOS projected to C and H atoms.} \label{band}
\end{figure}

Graphane, in its chair conformation as illustrated in
Fig.~\ref{band}(a), is derived by the adsorption of a single
hydrogen atom to each carbon atom alternating between the top
(\textit{A}) and bottom (\textit{B}) side in the honeycomb
structure. A charge of 0.1 electrons is transferred from H to C
leaving behind positively charged H atoms on both sides of a
double layer of negatively charged (-0.1 electrons) C atoms.
Graphane having a 2D quadruple structure has the work function
$\Phi$=4.97 eV, which is $\sim 0.2$ eV larger than that of graphene.
In contrast to semimetallic graphene, graphane is a semiconductor
with a wide direct band gap of 3.42 eV calculated by LDA but
corrected to be 5.97 eV with GW$_{0}$ self-energy method, as shown
in Fig.~\ref{band}(b). Doubly degenerate states at the
$\Gamma$-point at the top of the valence band are derived from
$2p_{x}$- and $2p_{y}$-orbitals of carbon atoms. The edge of the
conduction band is composed mainly from C-$p_{z}$ orbitals.
Calculated phonon bands all having positive frequencies confirm the
stability of 2D graphane. High frequency vibration modes associated
with C-H bonds are well separated from the rest of the spectrum, in
Fig.~\ref{band}(c).

The creation of a single H-vacancy at the hydrogen
covered surfaces gives rise to the spin polarization in the
non-magnetic perfect graphane. Desorption of a single H atom from
graphane is an endothermic reaction with 4.79 eV energy. Various
techniques, such as laser beam resonating with surface-hydrogen
bond\cite{laser}, stripping with ionic vapor\cite{breaux} and
scission of C-H bonds with subnanometer Pt clusters\cite{vajda},
can be used to create H-vacancy(ies). Upon desorption of a single
hydrogen atom, local bonding through $sp^{3}$ hybrid orbital is
retransformed into planar $sp^{2}$ and perpendicular $p_{z}$
($\pi$) orbitals. At the vacancy site one unpaired electron
accommodated by the dangling $p_{z}$ orbital contributes to the
magnetization by one $\mu_{B}$ (i.e. Bohr magneton). The exchange interaction between two H-vacancies calculated in a (11x11x1) supercell is found to be non-magnetic for the first and second nearest neighbor distances due to spin pairings. Since the $\pi$-$\pi$ interaction vanishes for farther distances, antiferromagnetic (AFM) state between two H-vacancies for the third and fourth nearest neigbor distance is energetically favorable. The occurrence of long range spin interactions in carbon based structures was explained before by the superexchange\cite{superexchange} and magnetic tail interaction\cite{tail}.

\begin{figure}
\includegraphics[width=8.5cm]{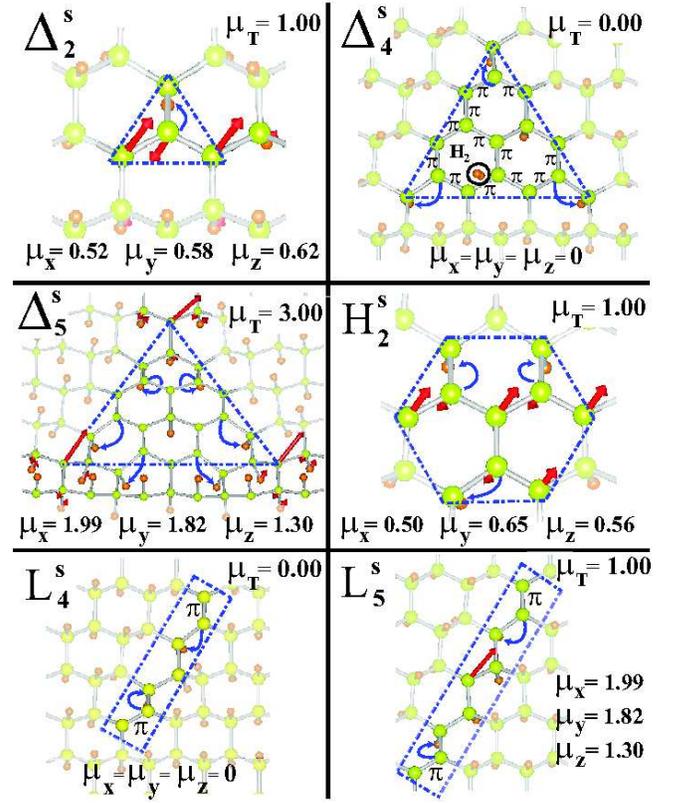}
\caption{(Color online) Calculated magnetic state of various
domains of single-sided H-vacancies, where all H atoms attached to
C atoms from upper side in the unshaded region (delineated by dash-dotted lines) including edges, are removed. The triangles  are
specified by $\Delta_{n}^{s}$ with $n$ indicating the maximum number of C atoms at one edge and $s$ signifies the single-sided
dehydrogenation. Similar symbols are used also for hexagonal,
$H_{2}^{s}$ and lane $L_{n}^{s}$ ($n=$4,5) domains. Total magnetic
moment $\mu_T$ and its components $\mu_{x}$, $\mu_{y}$ and
$\mu_{z}$ are given in units of the Bohr magneton $\mu_{B}$.
Magnetic moments on C atoms are shown by red (black) arrows.
Relocations of H atoms at the other side of graphane are shown by
curly arrows. For the sake of clarity $\pi$-bonds formed after the relocation of bottom H atoms are indicated only for $\Delta_{4}^{s}$, $L_{4}^{s}$ and $L_{5}^{s}$ structures.} \label{1sided}
\end{figure}

As for the islands of H-vacancies at the single (top) side of
graphane, we consider various geometrical domains, where H atoms
at their edges and inside are removed as seen in Fig.\ref{1sided}.
For a triangular domain specified as $\Delta_{2}^{s}$ at the top
side, H atoms attached to three carbon atoms located at each edge
are removed. Hydrogen atom which is normally adsorbed on the
central C atom at the bottom side moves to the corner. Under these
circumstances, spins of three hydrogen-free C atoms are
antiferromagnetically ordered to yield a net magnetic moment of 1
$\mu_{B}$. Noncollinear calculations with spin-orbit interaction
fix the directions of spins, which are tilted relative to the
normal to the graphane plane. For $\Delta_{4}^{s}$, a triangular
domain has ten H atoms removed from the top side of graphane.
While part of six H atoms are attached to carbon atoms from bottom
are relocated, remaining two H atoms are released by forming
H$_{2}$ molecule. At the end spins are paired and the net magnetic
moment of the domain becomes vanished. Generally, for a small single-sided
domain, $\mu_T$=0 if $N_t$, the total number of H atoms stripped,
is an even number so that adjacent $\pi$-orbitals form spin paired $\pi$-bonds. In this case, H atoms below the domain are
relocated (without facing any energy barrier) to pair adjacent $\pi$-orbitals to form maximum number
of $\pi$-bonds. At the end, a large buckled regions inside the
domain tends to be flattened and reconstructed to make nonmagnetic
graphene-like planar structure. In the case of $\Delta_{5}^{s}$,
while spins are paired through the formation of $\pi$-bonding
between two adjacent C atoms following the relocation H atoms at
the bottom side, the unpaired spins at the corner atoms  are
aligned in the same direction to yield a net magnetic moment of
$\mu_T$=3 $\mu_B$. The tendency to pair the spins of adjacent C
atoms to form $\pi$-bonds are seen better in lane domains. Let us
consider $L_{4}^{s}$ and $L_{5}^{s}$ in Fig.\ref{1sided}. Because
of relocation of H atoms at the bottom side, two pairs of nearest
neighbor C atoms form $\pi$-bonds and hence pair their spins. At
the end, $L_{4}^{s}$ has $\mu_{T}$=0. For $L_{5}^{s}$ having odd
number of H-vacancy, while two pairs of C atoms are bound by two
$\pi$-bonds, C atom at the center has an unpaired spin and attains
$\mu_{T}$=1 $\mu_{B}$. In a similar manner, the hexagonal domain
$H_{2}^{s}$ has total of seven C atoms at its center and corners,
all H atoms stripped from top side. At the bottom side, H atoms
are relocated and hence the spins of adjacent C atoms are paired
to result in a total net magnetic moment of $\mu$=1 $\mu_B$.

\begin{figure}
\includegraphics[width=8.5cm]{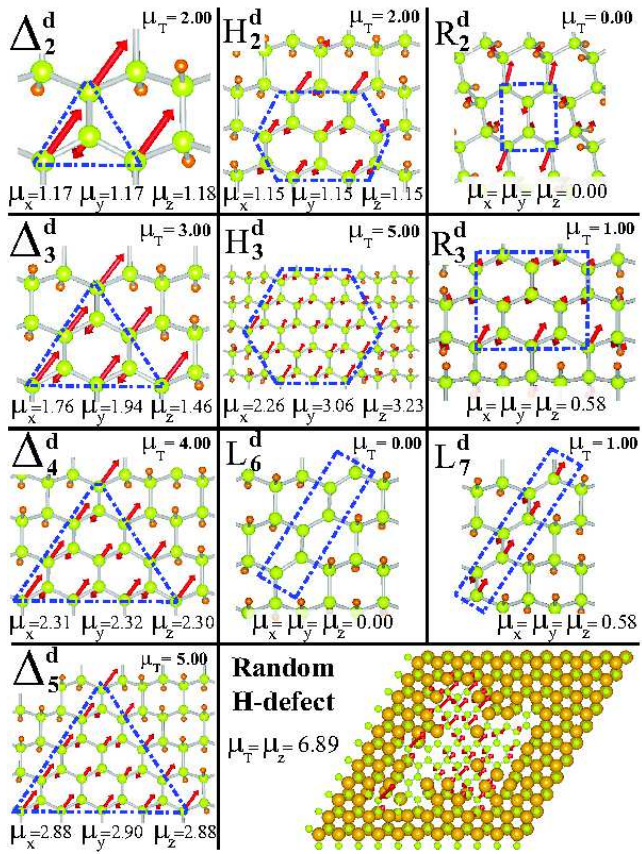}
\caption{(Color online) Net magnetic moments in Bohr magneton
within the triangular $\Delta_{n}^{d}$, hexagonal $H_{n}^{d}$,
rectangular $R_{n}^{d}$ and lane $L_{n}^{d}$ domains, which are
delineated by dash-dotted lines and have $n$ carbon atoms at their
edges. Here $d$ signifies the double-sided dehydrogenation. Random shaped domain including both one and two-sided H-vacancy parts is also illustrated.}
\label{2sided}
\end{figure}

We next show in Fig.\ref{2sided} that the magnetic moment of
graphane can be tuned by changing the size and geometry of a given
double-sided H-vacancy domain. In this case the situation is not
complex and allows us to figure out the magnetic moment of the
entire structure easily. Based on noncollinear calculations
including the spin-orbit coupling, the direction of the unpaired
spins on the \textit{A}-type C atoms freed from H atoms is found to
be opposite to that of the spins of \textit{B}-type C atoms.
However, instead of AFM spin ordering, lowest energy state of lane
defects consisting of even number of C atoms is NM due to the
entirely paired $p_{z}$ orbitals. Also, large double-sided domains
including lane defects with equal number of \textit{A}- and
\textit{B}-type C atoms are found to be NM. The resulting net
magnetic moment of a double-sided H-vacancy domains can be given by
$\mu_{T} = (N_{t}-N_{b})\mu_{B}$, where $N_{t}$ and $N_{b}$ denote
the number of stripped H atoms from the top and bottom sides,
respectively. Accordingly, the net magnetic moment
induced in $\Delta_{2}^{d}$, $\Delta_{3}^{d}$, $\Delta_{4}^{d}$ and
$\Delta_{5}^{d}$ domains are 2, 3, 4 and 5 the $\mu_{B}$
respectively. The same argument can be applied to rectangular
\textbf{$R_{n}^{d}$}, hexagonal \textbf{$H_{n}^{d}$} and lane
\textbf{$L_{n}^{d}$} domains. Even the magnetic moment of a domain
having arbitrary shape including various single-sided and
double-sided H-vacancy parts, as indicated in Fig.\ref{2sided}, can
be retrieved by the arguments discussed above. Non-integer value of
$\mu_{T}$ is due to severe distortion of structure. We also note
that our results regarding to the unpaired spin of a domain and
their net magnetic moment are in compliance with Lieb's
theorem\cite{lieb}, which distinguishes \textit{A}- and
\textit{B}-sublattices in honeycomb structure.

In conclusion, we showed that the interaction between unpaired spins
associated with H vacancies in graphane gives rise to interesting
magnetic structures. We revealed simple physical mechanisms
underlying the magnetism of single-sided and double-sided vacancy
domains. For single-sided domains, owing to the tendency to pair
the spins of $\pi$-orbitals of adjacent C atoms, some of the
adsorbed H atoms at the bottom side are relocated. At the end, the
net magnetic moments can be attained in vacancy domains depending
on their size and shape. For double-sided domains, interactions
underlying the generation of net magnetic moment are relatively
straightforward and are in good agreement with Lieb's theorem.
Since the exchange coupling between different domains are hindered
by domain walls, very dense data storage can be achieved through
uniform coverage of identical domains. It is also noted that a
graphane flake comprising a domain with large magnetic moment can
be utilized as a non-toxic marker for imaging purposes. While
magnetic 2D systems attract a great deal of attention due to their
tunable properties at nanoscale, our results suggest that the size
and ordering of magnetic moments of hydrogen vacancy domains with
thin walls can be used for future data storage and spintronics
applications.

Computing resources used in this work were partly provided by the
National Center for High Performance Computing of Turkey (UYBHM)
under grant number 2-024-2007.

\end{document}